\documentclass[a4paper,11pt]{article}
\usepackage{pos}
\usepackage{csquotes}

\title{Science4Peace in difficult times}

\author*[a]{Pierre Van Mechelen}

\affiliation[a]{Particle Physics group, Antwerp University, \\
Groenenborgerlaan 171, 2020 Antwerp, Belgium \\
on behalf of the Science4Peace Forum}

\emailAdd{Pierre.VanMechelen@uantwerpen.be}

\abstract{The war on Ukraine has affected significantly scientific cooperation and communication in particle physics and also in many other fields of scientific, cultural and educational exchange. Immediately after the war in February 2022, many scientific institutions paused or banned scientific cooperation and exchange with Russian and Belorusian institutes and their scientists. Publications were out on hold or even banned, if Russian scientists were on the author list.

The Science4Peace (S4P) Forum was created in March 2022 as a consequence of these restrictions, as a completely independent Forum for open communication among scientists, using independent ZOOM rooms and webpages. The basic ideas of the S4P Forum are fully in line with the policy of IUPAP of supporting and encouraging free scientific exchange. In the course of discussions on the consequences of the war on Ukraine, the S4P Forum organized a high-level panel discussion on ``Sanctions in Science - One Year of Sanctions''.

The war on Ukraine has also increased enormously the risk of nuclear escalation. Together with 14 Nobel laureates and many Scientists, the S4P Forum launched an appeal to subscribe to the ``no-first use'' policy and urged the governments to subscribe to the Treaty of the Prohibition of Nuclear Weapons adopted by the United Nations.

The S4P Forum fully supports the ideas and activities originating from the International Year of Basic Sciences for Sustainable Development (IYBSSD) to open and to keep open discussion channels at all levels, using Science as a common language.}

\FullConference{The European Physical Society Conference on High Energy Physics (EPS-HEP2023)\\
 21-25 August 2023\\
Hamburg, Germany\\}

\begin{document}
\maketitle

\section{The war in Ukraine and scientific cooperation across borders}

The invasion of Ukraine has resulted in unspeakable atrocities and war crimes.  We condemn aggression in the strongest possible terms; no arguments can ever justify violence.  As scientists, we experience how these terrible events profoundly affect international scientific cooperation and we wish to raise the question of how the international scientific community should react.

Historically, CERN has been a leading example of scientific cooperation across borders.  The history of cooperation between CERN and Sovjet institutes between 1955 and 1970, e.g., can be found \cite{CERNJINRHistory}. 
In "CERN's Main Objectives for the period 2021-2025" \cite{CERNObjectives} this is formulated as follows:

\begin{displayquote}
Since it was founded in 1954, CERN has promoted scientific collaboration across borders, inclusiveness and open science, \emph{transcending political and other conflicts}. CERN’s values, which are enshrined in the Convention, are as relevant as ever. \emph{The “CERN model” of global cooperation is taken as an example by other intergovernmental organisations and endeavours.}
\end{displayquote}

\noindent In addition, the mission statement of CERN as described on its webpages \cite{CERNMission} outlines four objectives:

\begin{itemize}
\item perform world-class research in fundamental physics; 
\item provide a unique range of particle accelerator facilities that enable research at the forefront of human knowledge, in an environmentally responsible and sustainable way;
\item \emph{unite people from all over the world to push the frontiers of science and technology, for the benefit of all;}
\item train new generations of physicists, engineers and technicians, and engage all citizens in research and in the values of science.
\end{itemize}

How should these principles be uphold today?

\section{Sanctions on scientific cooperation}

As a reaction to the invasion of Ukraine in February 2022, most Western countries and funding agencies imposed sanctions such as ceasing funding of (new) collaborative research projects or end formal partnerships and exchanges at the level of institutes.  

At the same time, explicit statements were made to express the intention to not impose a blanket suspension of academic links and to ensure that a dialogue remains possible.   Many funding agencies want to support Ukrainian and Russian colleagues who are currently abroad and individual students and researchers are still welcome at Western universities.  It is generally acknowledged that Russian colleagues that do not support the invasion  may feel abandoned by too severe sanctions.  Therefore, the general objective of sanctions is to target the regime, not the individual scientists.

At the level of institutes, sanctions range from cautious (adhering to national guidelines) to bold.  E.g., DESY imposed additional sanctions \cite{DESYMeasures}: banning visitors affiliated to Russian institutes, restricting access to IT infrastructure by individuals affiliated to Russian institutes, banning publications with co-authors affiliated to Russian institutes, and denying participation to conferences by scientists affiliated to Russian institutes.

The CERN Council has suspended the observer status of the Russian Federation in the Council \cite{CR3626}.  It has also suspended the participation of CERN scientists in scientific committees in Russia and vice versa, cancelled the organisation of joint events, and put on hold the issuing of new contracts of association with CERN for individuals affiliated to Russian institutes \cite{CR3637-3638}.  The CERN council declared to have the intention to terminate CERN's International Cooperation Agreements (ICAs) with the Russian Federation and the Republic of Belarus at their expiration dates in 2024 (a final decision by the CERN Council is due for December 2023) and to review CERN's future cooperation with the Joint Institute for Nuclear Research well in advance of the expiration date of the current ICA in January 2025 \cite{CR3669-3670-3671}.  The measures taken by the CERN Council are in accordance with national guidelines issued in most member states, but have consequences for individuals working at CERN.

\section{Actions implemented by the LHC Collaborations}

Sanctions imposed by others prompted a yearlong discussion about authorship of Collaboration papers published by the LHC experiments.  This resulted in the following measures. Authors affiliated with Russian or Belorussian institutes, or with JINR, sign the Collaboration’s scientific publications with their names and ORCID identifiers (where available), and the institute affiliation is replaced, respectively, by the reference: “Affiliated with an institute [or an international laboratory] covered by a cooperation agreement with CERN".   The complete author list including all institute affiliations is made available to the journal in a non-public form for the purpose of machine-readable analysis or as historical data.  No acknowledgement to the Russian and Belarussian funding agencies and JINR is made. Finally, on request, the experiment management will release a certificate attesting the contribution of the aforementioned institutes and funding agencies, or of JINR, to the work presented in the publication.

One should note, however, that alternative solutions were  implemented by some non-LHC Collaborations.  E.g., the Belle II Collaboration lists the names of all individual authors without their affliations and country names.  Instead, individual authors are uniquely identified using their ORCID.  In addtion Belle II papers include a statement in the acknowledgements that accurately recognizes historical contributions pre-dating the Russian invasion of the Ukraine: "This work, based on data collected using the Belle II detector, which was built and commissioned prior to March 2019, was supported by ...(list of funding sources). These acknowledgements are not to be interpreted as an endorsement of any statement made by any of our institutes, funding agencies, governments, or their representatives."  Similar solutions were adopted by the Babar en Belle Collaborations. 

\section{The Science4Peace Forum}

The Science4Peace Forum \cite{S4P} was established shortly after the Russian invasion of Ukraine.  Its main motivations and purposes are to reflect on the consequences of the Russian invasion of Ukraine on the evolution of scientific cooperation, defend the principle that cooperation among scientists engaged in peaceful fundamental research, fully decoupled from military or other scopes, must be protected, and weigh into decision making processes.
There is no restriction for participation: anyone including individuals from Russia, Ukraine, etc. can join the Forum.

The Forum organises biweekly (online) meetings with discussions on topics such as scientific exchanges and cooperation across borders, the effect of sanctions and restrictions, publication strategies, and nuclear threats.  E.g., on April 12, 2023, a virtual panel discussion was held on the "One year of sanctions in science" \cite{S4PPanel, S4PDossier}, and on June 28, 2023 an online lecture was presented by Prof. T. Sauer on "The war in Ukraine" \cite{S4PSauer}.

The Science4Peace Forum launched several petitions.  A first petition, ``Stop the escalation spiral'' \cite{S4PPet1}, condemns the war against Ukraine and calls to remove sanctions against scientists, promote non-military and peaceful scientific projects, allow common scientific publications and participation in conferences, and avoid the escalation resulting from extending sanctions to scientific and personal relations within the physics community.

A second petition, ``No first use, never any use of nuclear weapons'' \cite{S4PPet2}, demands governments to subscribe to a no-first-use policy and was sent on the occasion of the G7 summit in Hiroshima of May 2023.  This petition was signed by 14 Nobel Laureates and many other scientists.  Some encouraging responses were received from countries like Canada and Switzerland. 

\section{Conclusion}

Cooperation between scientists has been a landmark in the progress of science and peace over the last several decades, providing a framework in which political disputes could be set aside for the benefit of scientific progress and its positive reflection on humanity.

The Russian invasion of Ukraine jeopardises these long-standing collaborations and has triggered severe sanctions imposed by western governments, funding agencies, and individual institutes.

Although many decision-making bodies thread carefully, using peaceful  scientific cooperation as a tool to impose sanctions is questionable. There is a need to hear the voice of scientists themselves.

The Science4Peace Forum provides a platform for such discussions.  To join the mailing list and be kept informed, please visit www.science4peace.com

\end{document}